\documentclass[12pt,a4paper,aps,tightenlines,nofootinbib,showkeys,superscriptaddress]{revtex4-1}
\usepackage{amssymb,amsfonts,amsmath,amstext,graphicx,hyperref}
\usepackage{tikz}
\usepackage{braket}
\usepackage{float}
\usepackage{graphicx}
\usepackage[normalem]{ulem}
\usepackage{amsmath}
\usepackage{natbib}
\usepackage{xcolor,colortbl}
\begin{document}
        \title{Violation of Bell Monogamy Relations}
        \author{Abhisek Panda}
\email{abhisek.panda@phy.iitb.ac.in}
\affiliation{Department of Physics, Indian Institute of Technology Bombay, Mumbai 400076, India}
\author{Chandan Datta}
\email{chandan@iiserkol.ac.in}
\affiliation{Department of Physics, Indian Institute of Technology Jodhpur, Jodhpur 342030, India}
\affiliation{Department of Physical Sciences, Indian Institute of Science Education and Research Kolkata, Mohanpur, West Bengal 741246, India}

        \author{Pankaj Agrawal}
\email{pankaj.agrawal@tcgcrest.org}
\affiliation{Centre for Quantum Engineering, Research and Education, TCG CREST, Kolkata, India}

        \date{\today}

        \begin{abstract}
\vspace{0.3in}

  The entangled multipartite systems, specially in pure states, exhibit the phenomenon of entanglement monogamy.
  Such systems also display the phenomenon of Bell nonlocality.
  Like entanglement monogamy relations, there are Bell monogamy relations. These relations suggest a sharing
  of nonlocality across the subsystems. The nonlocality, as characterized by Bell inequalities, of one subsystem
  limits the nonlocality exhibited by another subsystem. We show that the Bell monogamy relations can be violated
  by using local filtering operations. We consider permutation-symmetric multipartite pure states,
  in particular $W$ states, to demonstrate the violation. 
  
        \end{abstract}

        \maketitle
        \section{Introduction}

The entanglement is a quintessential quantum phenomenon that makes quantum resources more powerful than classical resources \cite{RevModPhys.81.865, nielsen2010quantum, PhysicsPhysiqueFizika.1.195}. This phenomenon has inspired many quantum communication protocols to transmit classical or quantum information from one party to another or multiple parties \cite{PhysRevLett.70.1895}. The entanglement also plays an important role in secure communication and quantum key generation \cite{PhysRevLett.67.661, RevModPhys.74.145}, and measurement-based quantum computation \cite{PhysRevLett.86.5188, PhysRevA.68.022312}. 
 Multipartite entangled states play a major role in many quantum communication protocols. Understanding the nature of this entanglement is important.
 In a multi-party scenario, there is a limit to the sharing of entanglement 
 between different parties. There are entanglement monogamy relations to support this limitation \cite{PhysRevA.61.052306, PhysRevLett.96.220503}.
 In an extreme scenario, if two particles are in a maximally entangled state, then neither of these particles can be entangled to any other particle. 
 The Bell nonlocality is a concept that is associated with entanglement in the quantum mechanical framework. A pure bipartite state always violates Bell-CHSH inequality \cite{PhysRevLett.23.880, PhysicsPhysiqueFizika.1.195}. Multipartite entangled states, especially pure states, also exhibit the phenomenon of Bell nonlocality. There can be multiple notions
 of nonlocality in the case of multipartite states \cite{GISIN1991201, PhysRevLett.74.2619, RevModPhys.86.419, PhysRevD.35.3066, PhysRevLett.65.1838}. However, as one would
 expect, there are monogamy relations for Bell-CHSH nonlocality. The amount of violation of the Bell-CHSH inequality by one subsystem of a multi-party system limits the amount of violation by other subsystems. For multipartite states, one can go beyond Bell-CHSH inequalities. For example, one can consider Mermin, Svetlichny, or minimal-scenario inequalities. One can introduce monogamy relations with respect to these inequalities.
 
 Bell-CHSH inequality-based monogamy relations are satisfied by multipartite states. One interesting feature of
these monogamy relations is that not all bipartite subsystems violate the  Bell-CHSH inequality. However, we show
that local operations can lead to a violation of these monogamy relations. We note that the Bell monogamy relations are satisfied by the subsystems. So any operation on the system state will not affect the viability of the monogamy relations.
If we apply operations on a multipartite state and get another multipartite state, this state will also respect Bell monogamy
relation. This is because any multipartite state respects the Bell monogamy relations.
This violation is easiest to see in the context of systems in permutation-symmetric states. The subsystems of a system in such a state have many interesting properties. In particular,  two-particle or three-particle subsystems have identical reduced density operators respectively, thus showing same behavior. We consider multipartite $W$ states and demonstrate the violation of Bell-CHSH monogamy relations. We also consider more general permutation-symmetric states. We also consider the monogamy for multipartite Bell inequalities. In analogy to Bell-CHSH monogamy relations, one can introduce Bell monogamy relations for multipartite Bell inequalities. We show that these relations are respected by multipartite permutation-symmetric states, but on using local filtering operations, these states violate Bell monogamy relations for multipartite Bell inequalities. We consider several multipartite Bell inequalities, and show this phenomenon.

The paper is organized as follows. In the next section, we introduce Bell
monogamy relations and discuss permutation-symmetric states. In the section III,
we consider violation of the Bell monogamy relations by three-qubit $W$ states. In the section IV, we generalize 
the discussion to $N$-qubit $W$ states. In the section V,
we generalize the notion of Bell monogamy relations beyond Bell-CHSH inequalities to multipartite Bell inequalities. In the final section, we have some conclusions. In the Appendix, we have a discussion of three-qubit permutation-symmetric states beyond  $W$ states.

\section{Bell-CHSH Monogamy Relations and Permutation-symmetric States}

In the case of multipartite systems, the entanglement between
    different subsystems cannot take arbitrary values. Given the entanglement between 
two subsystems, the amount of entanglement between other subsystems cannot be arbitrary but
will be limited. For example,
    consider a system of three qubits. If qubits $A$ and $B$ are maximally
    entangled, then they cannot entangle with a third qubit $C$. So, we
    will have a direct product state, like
$$ \ket{\Psi}_{ABC} = \ket{\phi^+}_{AB} \ket{\eta}_{C}, $$
where $\ket{\phi^+}_{AB}$ is a Bell state.

There is a trade-off of entanglement between various subsystems. 
This notion was formalized by Coffman-Wooters-Kundu \cite{PhysRevA.61.052306}, using the concurrence measure. They showed that for a three-qubit system $ABC$, if $C_{A|B}$ is the 
   concurrence of subsystem $AB$, and $C_{A|C}$ is the 
   concurrence of subsystem $AC$, and $C_{A|BC}$ is the concurrence of the bipartite subsystems $A$ and
   $BC$, then the following monogamy relation holds:

   \begin{equation} \label{eq1}
C^{2}_{A|B} +  C^{2}_{A|C} \le C^{2}_{A|BC}.
\end{equation}

This monogamy relation has been extended to $n$-qubit systems also \cite{PhysRevLett.96.220503}.
Not all measures of entanglement exhibit monogamy. However,
   for various measures, specific classes of states may show monogamy \cite{GISIN1991201}.
   
  Toner and Verstraete \cite{Toner:2006rvn} introduced the monogamy of Bell nonlocality \cite{PhysRevA.71.022101}. They considered a tripartite
  system $ABC$. For such a system, they introduced the relation:
  \begin{equation} \label{eq2}
      \langle \mathcal{B}_{AB}\rangle^2 + 
      \langle \mathcal{B}_{AC}\rangle^2 \leq 8,
  \end{equation}

where,
\begin{equation} \label{eq3}
\mathcal{B} = A_1 \otimes (B_1 + B_2) +  A_2 \otimes (B_1 - B_2).  
\end{equation}

Here $A_1, A_2$ and $B_1, B_2$ are dichomatic
observables that can take the values $\{-1,1\}$. We will refer to the quantity $\mathcal{B}$
as a Bell function. Subsequently, by considering arbitrary measurements
for each subsystem, Qin, Fei, and Li-Jost \cite{PhysRevA.92.062339} generalized 
the relation to

\begin{equation} \label{eq4}
      \langle \mathcal{B}_{AB}\rangle^2 + 
      \langle \mathcal{B}_{BC}\rangle^2 +
      \langle \mathcal{B}_{AC}\rangle^2  \leq 12.
  \end{equation}

The maximum value of the average of the quantity $\mathcal{B}$ can be $2\sqrt{2}$. This is Tsirelson's bound. 
However, the maximum local value of $\langle \mathcal{B} \rangle$ can only be $2$. From Eq.(\ref{eq4}), it
is clear that all bipartite subsystems of a multipartite system cannot violate the Bell-CHSH inequality. If one or two two-qubit subsystems violate the Bell-CHSH inequality, then the third subsystem cannot. This will be true even if all the subsystems are
entangled. The situation is stark when we consider a special class of states -- permutation-symmetric states. All two-qubit subsystems of such states have identical density operators. Therefore,  either the state of none of the subsystems violates the Bell-CHSH inequality, or all of them violate. But the violation of the Bell-CHSH inequality by all subsystems leads to the violation the Bell monogamy relations. This is what we observe.

       We consider a wide range of permutation-symmetric states. We shall see that none of their two-qubit subsystems violates
    the Bell-CHSH inequality, even if entangled. We first consider the simplest state - the $GHZ$ state. This state is a three-qubit permutation-symmetric state,
\begin{equation} \label{eq5}
\ket{GHZ}=\frac{1}{\sqrt{2}}(\ket{000}+\ket{111}).
\end{equation}
  Taking partial trace with respect to the third qubit,
\begin{equation}
     \rho_{12}     = {1 \over 2}  \ket{00}\bra{00}   +    {1 \over 2}  \ket{11}\bra{11}. 
\end{equation}

 This is a mixture of product states, so this state does not violate the Bell-CHSH inequality.
 Same is true for  $ \rho_{13}$ and  $\rho_{23} $.
Since none of the subsystems violate
 the Bell-CHSH inequality, the Bell monogamy relation is satisfied.
Same will be true for $n$-qubit Generalized GHZ state. These states respect Bell monogamy relations.


\section{Bell monogamy of three-qubit $W$ states}

In this section, we consider a three-qubit $W$ state as another example of permutation-symmetric state. The $W$ state is a three-qubit symmetric state that can be written as 
$$\ket{W}=\frac{1}{\sqrt{3}}(\ket{001}+\ket{010}+\ket{100}).$$ Since it is a
permutation-symmetric state, each two-qubit subsystems has the same density operator.  Taking the partial trace with respect to the 3rd qubit we get 
$$ \rho^{3}_{12}=\mathrm{Tr}_3(\ket{W}\bra{W})=\left(\begin{array}{llll}
\frac{1}{3} & 0 & 0 & 0 \\
0 & \frac{1}{3} & \frac{1}{3} & 0 \\
0 & \frac{1}{3} & \frac{1}{3} & 0 \\
0 & 0 & 0 & 0
\end{array}\right). $$

By using the Peres-Hordecki criterion \cite{Peres_criteria, Horodecki_criteria}, we can check if this state is entangled. We take the partial transpose and find the eigenvalues of this matrix.
The eigenvalues are: $\{ {1 \over 3}, {1 \over 3}, { 1 \pm \sqrt{5} \over 6} \}$.
The eigenvalue ${ 1 -  \sqrt{5} \over 6} $ is negative, so this state is entangled.
 (By symmetry, the other two-qubit subsystems are also entangled.) 
Does this state violate the Bell CHSH inequality? The density operator
$\rho^{3}_{12}$  has only two non-zero eigenvalues, one corresponding to the noise (the eigenvalue 1/3) and another corresponding to the entangled state (the eigenvalue 2/3). To see if $\rho^{3}_{12}$ shows Bell violation, we calculate the sum of the two largest eigenvalues of the matrix $\boldsymbol{U}=T^\dagger T$ and see if the sum is greater than 1 \cite{HORODECKI1995340}. Here the elements of matrix $T$ are given by $T_{ij}=\mathrm{Tr}(\sigma_i\otimes\sigma_j \rho_{12}^{3})$, where $i=\{1,2,3\}$ and $\sigma_i$s refer to Pauli matrices. The matrix $\boldsymbol{U}$ is found to be $$\boldsymbol{U}=\left(\begin{array}{lll}
\frac{4}{9} & 0 & 0 \\
0 & \frac{4}{9} & 0 \\
0 & 0 & \frac{1}{9}
\end{array}\right).$$
We see that the sum of the two largest eigenvalues is 8/9, which is not greater than 1, so this state does not violate the Bell-CHSH inequality. By symmetry, the same is true about the other two-qubit subsystems. Therefore, we see that all three two-qubit subsystems of the $W$ states are entangled, but don't violate the Bell-CHSH
inequality. So the Bell monogamy relations Eq.(\ref{eq2}) and Eq.(\ref{eq4}) are respected.

{\it Lemma}: There exist local operations on the state $\rho^{3}_{12}$, which lead to the state
that violates the Bell-CHSH inequality, and consequently the Bell monogamy relation.

{\it Proof}: Proof is by construction.
 The state $\rho_{12}^3$ is a mixture of a maximally entangled Bell state with 2/3 probability and single noise $\ket{00}\bra{00}$ with 1/3 probability.  For $\rho^{3}_{12}$, we can use local filter\cite{GISIN1996151,Hirsch_2016}
  \begin{equation}    F= h \ket{0}\bra{0}+\ket{1}\bra{1}.
  \end{equation}
  Here $h$ is a small real parameter. After the application of the same filter by both parties,  we  get 
  $$\Tilde{\rho}^{3}_{12}=\frac{(F\otimes F)\rho_{12}^{3}(F^\dagger\otimes F^\dagger)}{Tr\{(F\otimes F)\rho_{12}^{3}(F^\dagger\otimes F^\dagger)\}}=\left(\begin{array}{llll}
\frac{h^2}{2 + h^2} & 0 & 0 & 0 \\
0 & \frac{1}{2 + h^2} & \frac{1}{2 + h^2} & 0 \\
0 & \frac{1}{2 + h^2} & \frac{1}{2 + h^2} & 0 \\
0 & 0 & 0 & 0
\end{array}\right).$$ 
As $h<1$, we see that the noise is suppressed by $h^2$. We can calculate the matrix $\boldsymbol{U}=T^\dagger T$, where elements of $T$ are $T_{ij}=\mathrm{Tr}(\sigma_i\otimes\sigma_j \Tilde{\rho}^{3}_{12})$. The matrix $\boldsymbol{U}$ is $$\boldsymbol{U}=\left(\begin{array}{lll}
\frac{4}{(2+h^2)^2} & 0 & 0 \\
0 & \frac{4}{(2+h^2)^2} & 0 \\
0 & 0 & \frac{(2-h^2)^2}{(2+h^2)^2}
\end{array}\right).$$
Since $(2-h^2)^2\leq 2^2$, the two largest eigenvalues are identical eigenvalues, {\it i.e.} $4/(2 + h^2)^2$. The sum of the two largest eigenvalues is $8/(2 + h^2)^2$. So, for the Bell-CHSH violation, the condition is
\begin{align}
    &\frac{8}{(2+h^2)^2}\geq 1\nonumber\\
    \implies& h \leq \sqrt{2(\sqrt{2}-1)}.\nonumber
\end{align}
This shows that, for a sufficiently small $h$, the locally filtered state violates the Bell-CHSH inequality.
This leads to the violation of the Bell monogamy relations Eq.(\ref{eq2}) and Eq.(\ref{eq4}). This completes the proof.

\section{Bell monogamy of $N$-qubit $W$ states}
We now consider a $N$-qubit $W$ state and show that its two-qubit subsystems respect Bell monogamy relations. But on applying appropriate local filters, there is a violation of the Bell monogamy relation.
A general $N$ qubit $W$ state can be written as 
\begin{equation}
|W_N\rangle=\frac{1}{\sqrt{N}} \sum \operatorname{Perm}\{|00 \cdots 01\rangle\},
\end{equation}
where Perm\{\} represents all possible unique permutations. As we are interested in its two-qubit reduced state, we need to trace out all other $N-2$ qubits. On tracing out last $N-2$ qubits, we get
 
\begin{equation}
\begin{aligned}
\rho^{N}_{12} = {\mathrm Tr}_{N-2}[|W_N\rangle \langle W_N|]= & \frac{1}{N}\left[(|10\rangle+|01\rangle) (\langle10|+\langle01|) +|00\rangle \langle00| (N-2) \right] \\& =\left[\begin{array}{cccc}
\frac{N-2}{N} & 0 & 0 & 0 \\
0 & \frac{1}{N} & \frac{1}{N} & 0 \\
0 & \frac{1}{N} & \frac{1}{N} & 0 \\
0 & 0 & 0 & 0
\end{array}\right].
\end{aligned}
\end{equation}

To find if $\rho^{N}_{12}$ is entangled, we find the eigenvalues of its partial transpose.
The lowest eigenvalue is found to be $\frac{N-2-\sqrt{(N-2)^2+4} }{2 N}$, which is negative for all $N > 2$. The state $\rho^{N}_{12}$ can be described as a mixture of an entangled state with $2/N$ probability and $|00\rangle \langle 00|$ noise with $(N-2)/N$ probability. To check if $\rho^{N}_{12}$ violates Bell-CHSH inequality, we calculate the eigenvalues of matrix $\boldsymbol{U}=T^\dagger T$. The eigenvalues are: $\{\frac{4}{N^2}, \frac{4}{N^2}, \frac{\left(N-4\right)^2}{N^2} \} $. As the sum of largest two eigenvalues is never greater than 1, the state $\rho^{N}_{12}$ does not violate Bell's inequality. Therefore Bell monogamy relations are respected.

{\it Lemma}: There exist local operations on $\rho^{N}_{12}$ state, so that the resulting state
violates the Bell-CHSH inequality, and consequently the Bell monogamy relations.

{\it Proof}: Proof is by construction.
As there is only single perpendicular noise in the state $\rho^{N}_{12}$, we can use the filter $F=h\ket{0}\bra{0}+\ket{1}\bra{1}$ on both qubits, where $0\leq h\leq 1$. After applying the filter, we get \begin{equation}
    \Tilde{\rho}^{N}_{12} =\frac{F \otimes F (\rho^{N}_{12}) F^{\dagger}\otimes F^{\dagger}}{\mathrm{Tr}(F \otimes F (\rho^{N}_{12}) F^{\dagger}\otimes F^{\dagger})} =\left(
\begin{array}{cccc}
 \frac{h^2 (N-2)}{h^2 (N-2)+2} & 0 & 0 & 0 \\
 0 & \frac{1}{h^2 (N-2)+2} & \frac{1}{h^2 (N-2)+2} & 0 \\
 0 & \frac{1}{h^2 (N-2)+2} & \frac{1}{h^2 (N-2)+2} & 0 \\
 0 & 0 & 0 & 0 \\
\end{array}
\right)
\end{equation}
The sum of the largest two eigenvalues of matrix $\boldsymbol{U}$ for $\Tilde{\rho}^{N}_{12}$ is either $s_1 = \frac{8}{\left(h^2 (N-2)+2\right)^2}$ or $s_2 = \frac{\left(h^2 (N-2)-2\right)^2+4}{\left(h^2 (N-2)+2\right)^2} $.\\

Given a $N$, we can always choose a $h$ such that $ h^2(N-2)<4$. Then $s_1 > s_2$
 We will have the violation of Bell-CHSH inequality if $s_1 > 1$. This will happen if
$$ h < \sqrt{\frac{2(\sqrt{2}-1)}{N-2}}$$
So for a given $N$, one can find a $h$, so that the filtered state violates the Bell-CHSH inequality. This will lead to the
violation of Bell monogamy relations. This completes the proof.

\section{Bell monogamy beyond Bell-CHSH inequality}

\subsection{Four-qubit $W$ States}

Until now, we have discussed the Bell monogamy relation
for two-qubit subsystems. What about multi-qubit subsystems beyond
two qubits? We will first consider three-qubit subsystems of generalized four-qubit $W$ state. For such subsystems, we will
consider monogamy with respect to four different Bell-type inequalities - Mermin inequality \cite{PhysRevLett.65.1838}, Sevetlichny inequality \cite{PhysRevD.35.3066}, DDA
inequality \cite{DDA}, and minimal-scenario facet inequality \cite{DAS20173928}.

For a three-qubit state, the Mermin inequality is
\begin{equation}
    A_1 B_1 C_2 + A_1 B_2 C_1 + A_2 B_1 C_1 - A_2 B_2 C_2 \leq 2.\label{eq19}
\end{equation}
The Svetlichny inequality is
\begin{equation}
  A_1 (B_1+B_2) C_1 + A_1 (B_1-B_2) C_2 + A_2 (B_1-B_2) C_1 - A_2 (B_1+B_2) C_2 \leq 4.
  \label{eq20}
\end{equation}
We also consider DDA inequality 
\begin{align}
& A_1\left(B_1+B_2\right)+A_2\left(B_1-B_2\right) C_1 \leq 2. 
\end{align}
Along with these inequalities, we also consider the minimal scenario facet inequality, 
\begin{equation}
I_{\text{CHSH}} + I_{\text{CHSH}} C_1 - 2C_1 \leq 2,  \label{eq22}
\end{equation}
where $I_{\text{CHSH}} = (A_1\left(B_1+B_2\right)+A_2\left(B_1-B_2\right))$. In all these inequalities, it is understood that
one has to take the average of the Bell functions on the left-hand side of these inequalities.
In the DDA and minimal scenario facet inequalities, there is one measurement on one of the qubits, and two measurements each
on the other two qubits. In each case, depending on the qubit with one measurement, there will be three such inequalities. However, for a permutation-symmetric state, all three will give identical results.
So we consider only one of such inequalities.
In all these four inequalities, $A_1, A_2$ are measurement settings for the first qubit (qubit $A$), $B_1, B_2$ are measurement settings for the second qubit (qubit $B$), and $C_1, C_2$ are measurement settings for the third qubit (qubit $C$). All are dichomatic observables with outcomes $\{\pm 1 \}$.

Do we have a monogamy relation with respect to these inequalities? In analogy to Bell-CHSH monogamy relation Eq.(\ref{eq4}), one may propose the following monogamy relations:

\begin{equation}
      \langle \mathcal{B}_3^{ABC}\rangle^2 +
      \langle \mathcal{B}_3^{ABD}\rangle^2 +
      \langle \mathcal{B}_3^{ACD}\rangle^2   +
      \langle \mathcal{B}_3^{BCD}\rangle^2          \leq \mathcal{C}_3. \label{c3}
\end{equation}
 The constant $\mathcal{C}_3$ will depend on the inequality. For Mermin, Minimal scenario facet, and DDA, the constant will be 16. For the Svetlichney inequality,  $C_3$ will be 64. As for the Bell-CHSH monogamy relation, we see that all three-qubit subsystems cannot violate the multipartite Bell inequalities to respect the monogamy relation. As before, it means that
 for a multipartite permutation-symmetric state, none of the three-qubit subsystems will violate any of the three-qubit multipartite
 Bell inequalities. As we shall see, these Bell monogamy relations are respected by multipartite $W$ state.




 Let us consider a four-qubit $W$ state, 
 \begin{equation}
\ket{W_4}=\frac{1}{2}(\ket{0001}+\ket{0010}+\ket{0100}+\ket{1000}),     
 \end{equation} 
and trace-out one of the qubits. We get a mixed state, which can be treated as a noisy three-qubit $W$ state. The resulting three-qubit state can be written as 

 \begin{equation}
\rho^{4}_{3}=\frac{1}{4}\ket{000}\bra{000}+\frac{3}{4}\ket{W_3}\bra{W_3},
\end{equation}
where $\ket{W_3}\equiv \ket{W} = \frac{1}{\sqrt{3}}(\ket{001}+\ket{010}+\ket{100}$.

 

We now wish to find out if $\rho^{4}_{3}$ violates any of the above multipartite Bell inequalities. This can be checked
by numerically maximizing all the Bell functions given in left side of the inequalities in Eq.(\ref{eq19}-\ref{eq22}) over
all measurement settings.
The state $\rho^{4}_{3}$ does not violate any inequalities given in Eq.(\ref{eq19}-\ref{eq22}). The maximum values of the Bell functions
are given in Table I. We see that none of the above multipartite Bell inequalities is violated, and so multipartite
Bell monogamy relation Eq.(\ref{c3}) is respected.

\begin{table}
    \centering
    \begin{tabular}{|c|c|c|c|c|}
    \hline
      Inequalities   & DDA  & Minimal Scenario  & Svetlichny & Mermin \\\hline
         Max value & 1.41421  &2  &2.82843 &2\\\hline
    \end{tabular}
    \caption{
Maximum values of the Bell functions for various tripartite Bell inequalities after maximizing over all measurement settings
    }
\end{table}

Let us now see if on applying a local filtering operation to $\rho^{4}_{3}$, the resulting state violates any of the above
multipartite Bell inequalities or not.
Since the three-qubit state $\rho^{4}_{3}$ only has a single noise, {\it i.e.} the noise in $\ket{000}\bra{000}$, we can filter out this noise before checking the violation in Eq.(\ref{eq19}-\ref{eq22}). Since we only have noise in $\ket{0}\bra{0}$ of every qubit, we can use the same filter on each qubit, {\it i.e.}, $F_A=F_B=F_C=F=h\ket{0}\bra{0}+\ket{1}\bra{1}$. After applying the filter we get 
$$
\begin{aligned}
     \Tilde{\rho}^{4}_{3} = \frac{F^{\otimes 3}(\rho^{4}_{3})F^{\dagger\otimes 3}}{\mathrm{Tr}(F^{\otimes 3}(\rho^{4}_{3})F^{\dagger\otimes 3})}
                          =\frac{1}{3+h^2}(h^2\ket{000}\bra{000}+3\ket{W_3}\bra{W_3})\\
\end{aligned}.
$$
For various values of $h$, we numerically maximize all the Bell functions given in left side of the inequalities in Eq.(\ref{eq19}-\ref{eq22}) over all measurement settings. The values of the Bell functions for various values of $h$ are listed in Table II.

\begin{table}[H]
    \centering
    \begin{tabular}{|c|c|c|c|c|}
    \hline
         & DDA & Minimal Scenario & Svetlichny & Mermin  \\\hline
         $h$=1 & 1.41421  &2  &2.82843 &2\\\hline
         $h$=0.99 &1.42837 & 2 & 2.84965 & \cellcolor{green!50}2.01501 \\\hline
         $h$=0.91 & 1.54253 &\cellcolor{green!50} 2.00034 & 3.02106 & \cellcolor{green!50}2.13579 \\\hline
          $h$=0.55 &\cellcolor{green!50} 2.03364 & \cellcolor{green!50}2.637 & 3.77026 &\cellcolor{green!50} 2.65413  \\\hline
          $h=0.4$&\cellcolor{green!50} 2.19845 &\cellcolor{green!50} 2.84611 &\cellcolor{green!50} 4.02814 &\cellcolor{green!50} 2.79747 \\\hline
    \end{tabular}
    \caption{
Maximum values of Bell functions of inequalities after maximizing over all measurement settings. The cells in which the inequality is violated for a particular value of $h$ is marked in green.
    }
    \label{table2}
\end{table}
From the Table \ref{table2}, we observe that as we increase the amount of filtering,  we violate more and more inequalities. Since for
a small enough $h$, there is a violation of all four multipartite Bell inequalities, this will lead to the violation of multipartite
Bell monogamy relation for all inequalities. So again, before the local filtering operation, the Bell monogamy relation is respected, but after filtering, it is not.

\subsection{Beyond Four-qubit W States}



One can generalize the three-qubit Bell monogamy relations to higher number of qubits. These monogamy relations
will also be not be violated by the subsystems of a permutation-symmetric state. As above, if we consider a four-qubit
subsystem state of a five-qubit permutation-symmetric state, it would not violate any of the four-qubit Bell inequalities.
By considering a five-qubit $W$ state, we will show that this is true. Again we will see that local filtering operations
will lead to the violation of four-qubit Bell inequalities.

For four-qubits, we consider two sets of inequalities -- DDA inequalities \cite{DDA} and minimal-scenario facet inequalities \cite{DAS20173928}.
As before for a symmetric state, there is only one independent inequality in the each set.
The minimal-scenario facet inequalities for four qubits is \cite{DAS20173928}:

 \begin{equation}
\begin{aligned}
\left(-2+A_1\left(B_1+B_2\right)+A_2\left(B_1-B_2\right)\right)\left(1+C_1\right)\left(1+D_1\right) \leq  & 0, \; \;\;\mathrm{or} \\
I_{CHSH}(1 + C_1)(1 + D_1) - 2 (C_1 +  D_1 + C_1 D_1) \leq & 2,
\label{4qfacet}
\end{aligned}
\end{equation}
where $I_{CHSH}=A_1\left(B_1+B_2\right)+A_2\left(B_1-B_2\right).$ The DDA inequality for four qubits is given as
\begin{equation}
    A_1 B_1 C_1(D_1+D_2)+ A_2 B_2 C_2(D_1-D_2) \leq 2.
    \label{4qbell}
\end{equation}
In these inequalities, it is understood that one has to take average of the left-hand side Bell function.
Here, $A_i, B_i, C_i$, and $D_i$ are observables as discussed earlier.

Following is a proposed Bell monogamy relation for the four-qubit subsystems 
\begin{equation}
      \langle \mathcal{B}_4^{ABCD}\rangle^2 +
      \langle \mathcal{B}_4^{ABCE}\rangle^2 +
      \langle \mathcal{B}_4^{ABDE}\rangle^2   +
      \langle \mathcal{B}_4^{ACDE}\rangle^2 +
      \langle \mathcal{B}_4^{BCDE}\rangle^2  \leq \mathcal{C}_4. \label{c4}
\end{equation}

For the above two Bell inequalities, $\mathcal{C}_4$ is $20$. As before, we can obtain a four-qubit reduced state 
from a five-qubit $W$ state. The four-qubit state is:

\begin{equation}
    \rho^{5}_{4} = \mathrm{Tr}_1(\ket{W_5}\bra{W_5})=\frac{1}{5}\ket{0000}\bra{0000}+\frac{4}{5} \ket{W_4}\bra{W_4}.
\end{equation}

Now, we numerically maximize over all the measurement settings to get the maximum value of the Bell functions given in left side of Eq.(\ref{4qfacet}) and Eq.(\ref{4qbell}). 
We observe that the state $\rho^{5}_{4}$ does not violate the minimal-scenario facet inequality and the DDA inequality. So, corresponding Bell
monogamy relations are respected. However, if we use the filter $F=h\ket{0}\bra{0}+\ket{1}\bra{1}$ for all the qubits as the noise is present only in $\ket{0000}\bra{0000}$ basis. After filtering, we will get 
\begin{equation}
    \Tilde{\rho}^{5}_{4} = \frac{F^{\otimes 4}(\rho^{5}_{4})F^{\dagger\otimes 4}}{\mathrm{Tr}(F^{\otimes 4}(\rho^{5}_{4})F^{\dagger\otimes 4})}=\frac{1}{4+h^2}(h^2\ket{0000}\bra{0000}+4\ket{W_4}\bra{W_4}).
\end{equation}
We again maximize over all the measurement settings to get the maximum value of the Bell functions given in left side of Eq.(\ref{4qfacet}) and Eq.(\ref{4qbell}). We observe that for $h\leq0.59$ DDA inequality is violated, and for $h\leq0.91$, the minimal-scenario facet inequality is violated. This means that corresponding Bell monogamy relations are violated.

\section{Arbitrary subsystems of $N$-qubit $W$ states}

As a further generalization, we can consider $M$-qubit subsystems of a $N$-qubit system. We can consider $M$-qubit
Bell inequalities, and corresponding monogamy relations. Again, we expect that if $N$-qubit system state is 
a permutation-symmetric state, then any $M$-qubit subsystem state will not violate a Bell Inequality to respect
the corresponding Bell monogamy relation. We find this to be true by considering a $N$-qubit $W$ state. Again,
we shall see that a local filtering operation leads to the violation of a Bell inequality, and thus the violation
of corresponding Bell monogamy relation. Here we will consider minimal-scenario facet Bell inequalities \cite{DAS20173928}.

We can generalize the Bell monogamy relation of Eq.(\ref{eq4}) as
\begin{equation}
   \sum_{(i₁,...,i_M)}   \langle \mathcal{B}_M^{(i₁,...,i_M)}\rangle^2          \leq \mathcal{C}_M, \label{cm}
\end{equation}
where $\mathcal{C}_M = C^N_M l^2_{\mathrm{max}}$ and $l_{\mathrm{max}}$ is the maximum local value of the Bell function.
Here, the sum is over all the choices of selecting $M$-qubit subsystems from an $N$-qubit system state. There
will be $C^N_M$ terms in the sum.

A general $N$-qubit $W$ state can be written as 

\begin{equation}
|W_N \rangle=\frac{1}{\sqrt{N}} \sum \operatorname{Perm}\{|00 \cdots 01\rangle\},
\end{equation}
where Perm\{\} represents all possible unique permutations. 
Tracing out the last $N-3$ qubits gives us 
\begin{equation}
\rho^{N}_{3} = \frac{N-3}{N}\ket{000}\bra{000} + \frac{3}{N}\ket{W_3}\bra{W_3}.
\end{equation}

Similarly, we can find $M$-qubit reduced state from $ N$-qubit $W$ state. The reduced state is
\begin{equation}
\rho^{N}_{M} = \frac{N-M}{N}\ket{00...0}\bra{00...0}+\frac{M}{N}\ket{W_M}\bra{W_M},
\end{equation}
 where $\ket{W_M}$ is the $M$-qubit $W$ state.
We have checked for several $M$ values that this state does not violate the corresponding minimal scenario facet Bell inequalities.
This state is a mixture of a $M$-qubit $W$ state and noise $(\ket{0}\bra{0})^{\otimes M}$. Since there is only one noise, we can use filter $F=h\ket{0}\bra{0}+\ket{1}\bra{1}$ on every qubit. After applying filter we get
\begin{equation}
    \Tilde{\rho}^{N}_{M} = \frac{F^{\otimes M} (\rho) F^{\dagger\otimes M}}{\mathrm{Tr}(F^{\otimes M} (\rho) F^{\dagger\otimes M})} =
 \frac{h^2 (N-M)}{h^2 (N-M)+M}(\ket{0}\bra{0})^{\otimes M}+
 \frac{1}{h^2 (N-M)+M}\ket{W_M}\bra{W_M}.
\end{equation}
For $M=2$, we can derive analytically that any filter with $h<\sqrt{2(\sqrt{2}-1)/(N-2)}$ violates the Bell-CHSH inequality, which is a facet Bell inequality. 

\begin{figure}[H]
    \centering
   \begin{minipage}[t]{0.48\textwidth}
        \centering
        \includegraphics[width=1.1\textwidth]{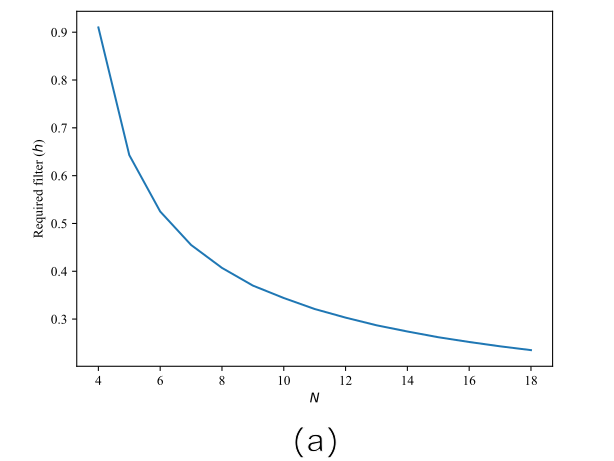}
        \label{fig:image1sub}
    \end{minipage}
    \begin{minipage}[t]{0.48\textwidth}
        \centering
        \includegraphics[width=1.1\textwidth]{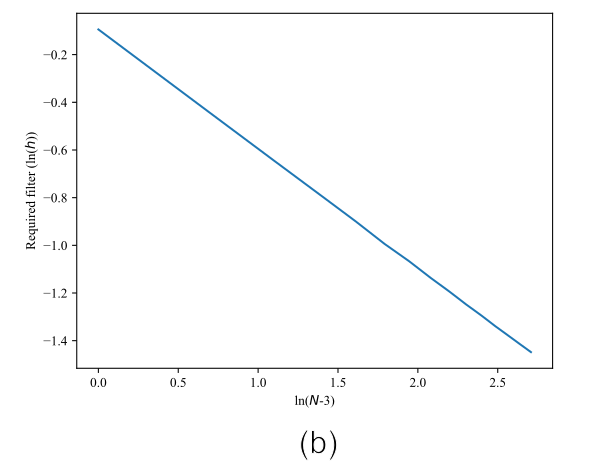}
        \label{fig:image2sub}
    \end{minipage}
    
    \caption{Maximum filter parameter ($h$) required for violation of Facet inequality of $M=3$ qubit reduced state from $N$ qubit $W$ state.}
    \label{fig:combinedfigure}
\end{figure}

For $M>2$, there is no analytical condition to check for the violation of a Bell inequality. We find the strength of the filter $(h)$, which leads to a violation of the facet inequality numerically. In Fig(\ref{fig:combinedfigure}),  we have plotted the maximum 
value of filter parameter $h$ for violation of the facet inequality for $M=3$ for various values of $N$.
 
We observe a straight line in the right panel of the Fig(\ref{fig:combinedfigure}), indicating that the maximum filter parameter is proportional to $1/\sqrt{N-3}$.
On repeating this process for $M=4$, we find that the maximum filter parameter is proportional to $1/\sqrt{N-4}$. 
A generalization to the higher qubit subsystems suggests that for the filter parameter $h<\sqrt{2(\sqrt{2}-1)/(N-M)}$, there may
be a violation of the corresponding minimal-scenario facet Bell inequality, and thus violation of the corresponding Bell monogamy relation.


\section{Discussion and Conclusions}

We have considered the phenomenon of Bell nonlocality for multipartite qubit states. Just like entanglement monogamy relations,
there are Bell monogamy relations. These relations suggest that for a system in a pure state, at least one of the subsystem state
does not show Bell nonlocality. For a permutation-symmetric state, none of the subsystem state exhibits Bell nonlocality.
These subsystem states respect the corresponding Bell monogamy relation. We have considered $N$-qubit $W$ states and showed 
that their two-qubit subsystems satisfy the Bell-CHSH monogamy relation. However, if we use a local filtering operation, then
these Bell monogamy relations are violated. We have generalized the Bell-CHSH monogamy relations to multi-qubit monogamy
relations. We show that these relations are satisfied with respect to multipartite Bell inequalities. However, local filtering
operations again violate the multipartite Bell monogamy relation. These investigations may help in understanding nonlocal properties
of multipartite states, and we may need a more robust set of Bell monogamy relations.

\bibliography{bibfie}

@article{DDA,
author = {Das, Arpan and Datta, Chandan and Agrawal, Pankaj},
title = {Minimal scenario facet Bell inequalities for multi-qubit states},
journal = {International Journal of Quantum Information},
volume = {21},
number = {01},
pages = {2350005},
year = {2023},
doi = {10.1142/S0219749923500053},
URL = {https://doi.org/10.1142/S0219749923500053},
eprint = {https://doi.org/10.1142/S0219749923500053},
}

@article{DAS20173928,
title = {New Bell inequalities for three-qubit pure states},
journal = {Physics Letters A},
volume = {381},
number = {47},
pages = {3928-3933},
year = {2017},
issn = {0375-9601},
doi = {https://doi.org/10.1016/j.physleta.2017.10.023},
url = {https://www.sciencedirect.com/science/article/pii/S0375960117310198},
author = {Arpan Das and Chandan Datta and Pankaj Agrawal},
}

@article{Horodecki_criteria,
title = {Separability of mixed states: necessary and sufficient conditions},
journal = {Physics Letters A},
volume = {223},
number = {1},
pages = {1-8},
year = {1996},
issn = {0375-9601},
doi = {https://doi.org/10.1016/S0375-9601(96)00706-2},
url = {https://www.sciencedirect.com/science/article/pii/S0375960196007062},
author = {M. Horodecki and P. Horodecki and Ryszard Horodecki}
}

@article{Peres_criteria,
  title = {Separability Criterion for Density Matrices},
  author = {Peres, Asher},
  journal = {Phys. Rev. Lett.},
  volume = {77},
  issue = {8},
  pages = {1413--1415},
  numpages = {0},
  year = {1996},
  month = {Aug},
  publisher = {American Physical Society},
  doi = {10.1103/PhysRevLett.77.1413},
  url = {https://link.aps.org/doi/10.1103/PhysRevLett.77.1413}
}

@article{RevModPhys.81.865,
  title = {Quantum entanglement},
  author = {Horodecki, Ryszard and Horodecki, Pawe\l{} and Horodecki, Micha\l{} and Horodecki, Karol},
  journal = {Rev. Mod. Phys.},
  volume = {81},
  issue = {2},
  pages = {865--942},
  numpages = {0},
  year = {2009},
  month = {Jun},
  publisher = {American Physical Society},
  doi = {10.1103/RevModPhys.81.865},
  url = {https://link.aps.org/doi/10.1103/RevModPhys.81.865}
}

@book{nielsen2010quantum,
  title={Quantum computation and quantum information},
  author={Nielsen, Michael A and Chuang, Isaac L},
  year={2010},
  publisher={Cambridge university press}
}

@article{PhysicsPhysiqueFizika.1.195,
  title = {On the Einstein Podolsky Rosen paradox},
  author = {Bell, J. S.},
  journal = {Physics Physique Fizika},
  volume = {1},
  issue = {3},
  pages = {195--200},
  numpages = {6},
  year = {1964},
  month = {Nov},
  publisher = {American Physical Society},
  doi = {10.1103/PhysicsPhysiqueFizika.1.195},
  url = {https://link.aps.org/doi/10.1103/PhysicsPhysiqueFizika.1.195}
}

@article{PhysRevLett.70.1895,
  title = {Teleporting an unknown quantum state via dual classical and Einstein-Podolsky-Rosen channels},
  author = {Bennett, Charles H. and Brassard, Gilles and Cr\'epeau, Claude and Jozsa, Richard and Peres, Asher and Wootters, William K.},
  journal = {Phys. Rev. Lett.},
  volume = {70},
  issue = {13},
  pages = {1895--1899},
  numpages = {0},
  year = {1993},
  month = {Mar},
  publisher = {American Physical Society},
  doi = {10.1103/PhysRevLett.70.1895},
  url = {https://link.aps.org/doi/10.1103/PhysRevLett.70.1895}
}

@article{PhysRevLett.67.661,
  title = {Quantum cryptography based on Bell's theorem},
  author = {Ekert, Artur K.},
  journal = {Phys. Rev. Lett.},
  volume = {67},
  issue = {6},
  pages = {661--663},
  numpages = {0},
  year = {1991},
  month = {Aug},
  publisher = {American Physical Society},
  doi = {10.1103/PhysRevLett.67.661},
  url = {https://link.aps.org/doi/10.1103/PhysRevLett.67.661}
}

@article{RevModPhys.74.145,
  title = {Quantum cryptography},
  author = {Gisin, Nicolas and Ribordy, Gr\'egoire and Tittel, Wolfgang and Zbinden, Hugo},
  journal = {Rev. Mod. Phys.},
  volume = {74},
  issue = {1},
  pages = {145--195},
  numpages = {0},
  year = {2002},
  month = {Mar},
  publisher = {American Physical Society},
  doi = {10.1103/RevModPhys.74.145},
  url = {https://link.aps.org/doi/10.1103/RevModPhys.74.145}
}

@article{PhysRevLett.86.5188,
  title = {A One-Way Quantum Computer},
  author = {Raussendorf, Robert and Briegel, Hans J.},
  journal = {Phys. Rev. Lett.},
  volume = {86},
  issue = {22},
  pages = {5188--5191},
  numpages = {0},
  year = {2001},
  month = {May},
  publisher = {American Physical Society},
  doi = {10.1103/PhysRevLett.86.5188},
  url = {https://link.aps.org/doi/10.1103/PhysRevLett.86.5188}
}

@article{PhysRevA.68.022312,
  title = {Measurement-based quantum computation on cluster states},
  author = {Raussendorf, Robert and Browne, Daniel E. and Briegel, Hans J.},
  journal = {Phys. Rev. A},
  volume = {68},
  issue = {2},
  pages = {022312},
  numpages = {32},
  year = {2003},
  month = {Aug},
  publisher = {American Physical Society},
  doi = {10.1103/PhysRevA.68.022312},
  url = {https://link.aps.org/doi/10.1103/PhysRevA.68.022312}
}

@article{PhysRevA.61.052306,
  title = {Distributed entanglement},
  author = {Coffman, Valerie and Kundu, Joydip and Wootters, William K.},
  journal = {Phys. Rev. A},
  volume = {61},
  issue = {5},
  pages = {052306},
  numpages = {5},
  year = {2000},
  month = {Apr},
  publisher = {American Physical Society},
  doi = {10.1103/PhysRevA.61.052306},
  url = {https://link.aps.org/doi/10.1103/PhysRevA.61.052306}
}

@article{PhysRevLett.96.220503,
  title = {General Monogamy Inequality for Bipartite Qubit Entanglement},
  author = {Osborne, Tobias J. and Verstraete, Frank},
  journal = {Phys. Rev. Lett.},
  volume = {96},
  issue = {22},
  pages = {220503},
  numpages = {4},
  year = {2006},
  month = {Jun},
  publisher = {American Physical Society},
  doi = {10.1103/PhysRevLett.96.220503},
  url = {https://link.aps.org/doi/10.1103/PhysRevLett.96.220503}
}

@article{PhysRevLett.23.880,
  title = {Proposed Experiment to Test Local Hidden-Variable Theories},
  author = {Clauser, John F. and Horne, Michael A. and Shimony, Abner and Holt, Richard A.},
  journal = {Phys. Rev. Lett.},
  volume = {23},
  issue = {15},
  pages = {880--884},
  numpages = {0},
  year = {1969},
  month = {Oct},
  publisher = {American Physical Society},
  doi = {10.1103/PhysRevLett.23.880},
  url = {https://link.aps.org/doi/10.1103/PhysRevLett.23.880}
}

@article{GISIN1991201,
title = {Bell's inequality holds for all non-product states},
journal = {Physics Letters A},
volume = {154},
number = {5},
pages = {201-202},
year = {1991},
issn = {0375-9601},
doi = {https://doi.org/10.1016/0375-9601(91)90805-I},
url = {https://www.sciencedirect.com/science/article/pii/037596019190805I},
author = {N. Gisin}
}

@article{PhysRevLett.74.2619,
  title = {Bell's Inequalities and Density Matrices: Revealing ``Hidden'' Nonlocality},
  author = {Popescu, Sandu},
  journal = {Phys. Rev. Lett.},
  volume = {74},
  issue = {14},
  pages = {2619--2622},
  numpages = {0},
  year = {1995},
  month = {Apr},
  publisher = {American Physical Society},
  doi = {10.1103/PhysRevLett.74.2619},
  url = {https://link.aps.org/doi/10.1103/PhysRevLett.74.2619}
}

@article{RevModPhys.86.419,
  title = {Bell nonlocality},
  author = {Brunner, Nicolas and Cavalcanti, Daniel and Pironio, Stefano and Scarani, Valerio and Wehner, Stephanie},
  journal = {Rev. Mod. Phys.},
  volume = {86},
  issue = {2},
  pages = {419--478},
  numpages = {60},
  year = {2014},
  month = {Apr},
  publisher = {American Physical Society},
  doi = {10.1103/RevModPhys.86.419},
  url = {https://link.aps.org/doi/10.1103/RevModPhys.86.419}
}

@article{PhysRevD.35.3066,
  title = {Distinguishing three-body from two-body nonseparability by a Bell-type inequality},
  author = {Svetlichny, George},
  journal = {Phys. Rev. D},
  volume = {35},
  issue = {10},
  pages = {3066--3069},
  numpages = {0},
  year = {1987},
  month = {May},
  publisher = {American Physical Society},
  doi = {10.1103/PhysRevD.35.3066},
  url = {https://link.aps.org/doi/10.1103/PhysRevD.35.3066}
}

@article{PhysRevLett.65.1838,
  title = {Extreme quantum entanglement in a superposition of macroscopically distinct states},
  author = {Mermin, N. David},
  journal = {Phys. Rev. Lett.},
  volume = {65},
  issue = {15},
  pages = {1838--1840},
  numpages = {0},
  year = {1990},
  month = {Oct},
  publisher = {American Physical Society},
  doi = {10.1103/PhysRevLett.65.1838},
  url = {https://link.aps.org/doi/10.1103/PhysRevLett.65.1838}
}

@article{Toner:2006rvn,
    author = "Toner, Benjamin and Verstraete, Frank",
    title = "{Monogamy of Bell correlations and Tsirelson's bound}",
    eprint = "quant-ph/0611001",
    archivePrefix = "arXiv",
    month = "11",
    year = "2006"
}

@article{PhysRevA.92.062339,
  title = {Trade-off relations of Bell violations among pairwise qubit systems},
  author = {Qin, Hui-Hui and Fei, Shao-Ming and Li-Jost, Xianqing},
  journal = {Phys. Rev. A},
  volume = {92},
  issue = {6},
  pages = {062339},
  numpages = {4},
  year = {2015},
  month = {Dec},
  publisher = {American Physical Society},
  doi = {10.1103/PhysRevA.92.062339},
  url = {https://link.aps.org/doi/10.1103/PhysRevA.92.062339}
}

@article{PhysRevA.71.022101,
  title = {Nonlocal correlations as an information-theoretic resource},
  author = {Barrett, Jonathan and Linden, Noah and Massar, Serge and Pironio, Stefano and Popescu, Sandu and Roberts, David},
  journal = {Phys. Rev. A},
  volume = {71},
  issue = {2},
  pages = {022101},
  numpages = {11},
  year = {2005},
  month = {Feb},
  publisher = {American Physical Society},
  doi = {10.1103/PhysRevA.71.022101},
  url = {https://link.aps.org/doi/10.1103/PhysRevA.71.022101}
}

@article{acinstate,
  title = {Generalized Schmidt Decomposition and Classification of Three-Quantum-Bit States},
  author = {Ac\'{\i}n, A. and Andrianov, A. and Costa, L. and Jan\'e, E. and Latorre, J. I. and Tarrach, R.},
  journal = {Phys. Rev. Lett.},
  volume = {85},
  issue = {7},
  pages = {1560--1563},
  numpages = {0},
  year = {2000},
  month = {Aug},
  publisher = {American Physical Society},
  doi = {10.1103/PhysRevLett.85.1560},
  url = {https://link.aps.org/doi/10.1103/PhysRevLett.85.1560}
}

@article{HORODECKI1995340,
title = {Violating Bell inequality by mixed spin-1/2 states: necessary and sufficient condition},
journal = {Physics Letters A},
volume = {200},
number = {5},
pages = {340-344},
year = {1995},
issn = {0375-9601},
doi = {https://doi.org/10.1016/0375-9601(95)00214-N},
url = {https://www.sciencedirect.com/science/article/pii/037596019500214N},
author = {R. Horodecki and P. Horodecki and M. Horodecki},
}

@article{Hirsch_2016,
doi = {10.1088/1367-2630/18/11/113019},
url = {https://doi.org/10.1088/1367-2630/18/11/113019},
year = {2016},
month = {nov},
publisher = {IOP Publishing},
volume = {18},
number = {11},
pages = {113019},
author = {Hirsch, Flavien and Quintino, Marco Túlio and Bowles, Joseph and Vértesi, Tamás and Brunner, Nicolas},
title = {Entanglement without hidden nonlocality},
journal = {New Journal of Physics},
}

@article{GISIN1996151,
title = {Hidden quantum nonlocality revealed by local filters},
journal = {Physics Letters A},
volume = {210},
number = {3},
pages = {151-156},
year = {1996},
issn = {0375-9601},
doi = {https://doi.org/10.1016/S0375-9601(96)80001-6},
url = {https://www.sciencedirect.com/science/article/pii/S0375960196800016},
author = {N. Gisin},
}
\vspace{0.5in}
 {\bf \Large Appendix A}   

\vspace{0.05in}



\vspace{0.1in}

In this appendix, we consider the Bell violation for a more general symmetric state. This state is
\begin{equation}
  \ket{\psi_{sym}}=a_1\ket{000}+b_1\ket{111}+ c_1 (\ket{001}+\ket{010}+\ket{100}),  \label{ap1}
\end{equation}
where $c_1 = \frac{\sqrt{1-a_1^2-b_1^2}}{\sqrt{3}} $.
To verify the Bell-CHSH monogamy relation, we find two-qubit subsystem density operator $\rho_{12}$
and compute the matrix $\boldsymbol{U}$ to check the Bell-CHSH violation. For the subsystem `12', the
density operator is

\begin{equation}
    \rho_{12}=\left(
\begin{array}{cccc}
 c_1^2+a_1^2 & a_1 c_1 & a_1 c_1 & b_1 c_1 \\
 a_1 c_1 & c_1^2 & c_1^2 & 0 \\
 a_1 c_1 & c_1^2 & c_1^2 & 0 \\
 b_1 c_1 & 0 & 0 & b_1^2 \\
\end{array}                             
\right),                                    \label{ap2}
\end{equation}
and compute the matrix $\boldsymbol{U}$ to check the Bell violation.
We find 3 eigenvalues of the matrix $\boldsymbol{U}$ and numerically maximize the sum of the two largest eigenvalues in the $0\leq (a_1,b_1) \leq 1$ range. The maximum of the sum of the two largest eigenvalues of $\boldsymbol{U}$ is 1 and this value is attained for $a_1=0.1968$ and $b_1=0.4902$. So the two-qubit state $\rho_{12}$ does not violate Bell-CHSH inequality. Same will be true
for $\rho_{23}$ and $\rho_{31}$. So we find that none of the two-qubit subsystems violate the Bell-CHSH inequalities, and the Bell monogamy relation Eq.($\ref{eq4}$) is respected.

To see what happens after applying local filters, let us find the state with two different
local filter operations. 
If one needs to suppress the $|00 \rangle \langle 00|$ component of the noise, one
can use the local filter $F_1=h\ket{0}\bra{0}+\ket{1}\bra{1}$. On applying
this filter, we get

\begin{equation}
\Tilde{\rho}^{1}_{12}=\frac{1}{N_1}\left(
\begin{array}{cccc}
 h^4 \left(2 a_1^2-b_1^2+1\right) & 3 a_1 h^3 c_1 & 3 a_1 h^3 c_1 & 3 b_1 h^2 c_1 \\
 3 a_1 h^3 c_1 &  3 h^2 c_1^2 & 3 h^2 c_1^2 & 0 \\
 3 a_1 h^3 c_1 & 3 h^2 c_1^2 & 3 h^2 c_1^2 & 0 \\
 3 b_1 h^2 c_1 & 0 & 0 & 3 b_1^2 \\
\end{array}
\right),                                \label{ap3}              
\end{equation}
where $N_1= h^2 \left(2 a_1^2 \left(h^2-1\right)+h^2+2\right)-b_1^2 \left(h^4+2 h^2-3\right)$. 

If one needs to suppress the $|11 \rangle \langle 11|$ component of the noise, one
can use the local filter $F_2= \ket{0}\bra{0} + h \ket{1}\bra{1}$, we get

\begin{equation}
\hspace{-2cm}\Tilde{\rho}^{2}_{12}=\frac{1}{N_2}\left(
\begin{array}{cccc}
 -2 a_1^2+b_1^2-1 & - 3 a_1 h c_1 & - 3 a_1 h c_1 & - 3 b_1 h^2 c_1 \\
 -3 a_1 h c_1 & 3 h^2 c_1^2 & 3 h^2 c_1^2 & 0 \\
 -3 a_1 h c_1 & 3 h^2 c_1^2 & 3 h^2 c_1^2 & 0 \\
 -3 b_1 h^2 c_1 & 0 & 0 & -3 b_1^2 h^4 \\
\end{array}
\right),                          \label{ap4}
\end{equation}
where $N_2 =2 a_1^2 \left(h^2-1\right)+b_1^2 \left(-3 h^4+2 h^2+1\right)-2 h^2-1$.


There are four possible situations -- (I) $b_1=0$ and $a_1 = 0$, (II) $b_1=0$ and $a_1 \neq 0$, (III) $a_1=0$ and $b_1 \neq 0$,
(IV) $b_1 \neq 0$ and $a_1 \neq 0$. Let us consider all cases, one by one.

\begin{itemize}

\item Case I:  We have ($b_1=0$ and $a_1 = 0$) 

In this case, the state $\ket{\psi_{sym}}$ reduces to a three-qubit $W$ state. This case has been
discussed in the main text. In this case, we have seen the violation of the Bell-CHSH monogamy relation
after a local filtering operation.

\item Case II: We have ($b_1=0$ and $a_1\neq 0$) 

If $b_1=0$ and $a_1 \neq 0$ then from the density operator Eq.(\ref{ap1}), we see that there is noise
in $|00 \rangle \langle 00|$ component. So we use the filter $F_1$, and the corresponding filtered 
state is given in Eq.(\ref{ap3}).
The sum of the two largest eigenvalues of matrix $\boldsymbol{U}$ for $\Tilde{\rho}^{1}_{12}$ is plotted by
varying values of $a_1$ and $h$ in the Fig.~(\ref{fig:apfig1}). We can see that there is Bell violation throughout
the range of $a_1$, and the violation increases as the strength
of the filter increases (lowering the value of $h$). This suggests that the Bell monogamy relation Eq.(\ref{eq4}) is violated
for all possible values of $a_1$.
\begin{figure}[H]
    \centering
    \includegraphics[width=0.75\linewidth]{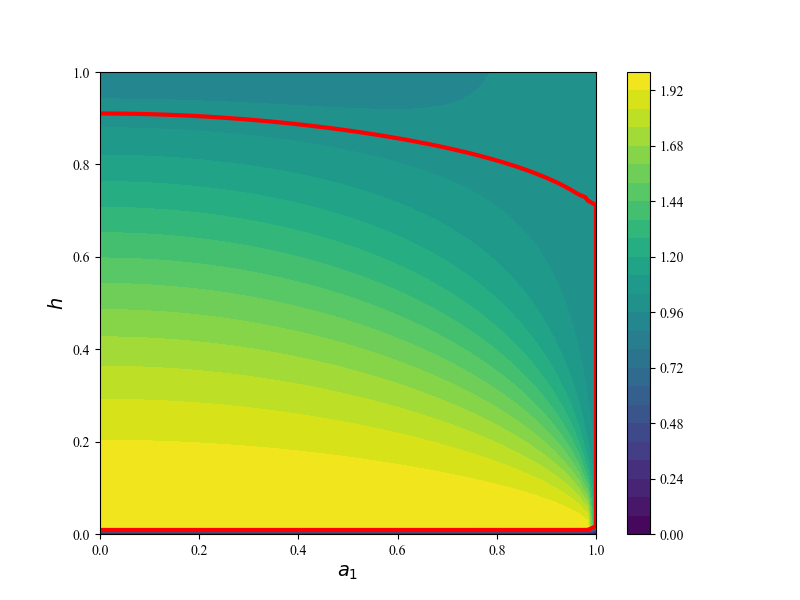}
    \caption{Contour plot of the sum of two largest eigenvalues of matrix $\boldsymbol{U}$ for $\Tilde{\rho}^{1}_{12}$,
    as a function of filter parameter ($h$) and the state parameter $a_1$. Here $b_1 = 0$ and the filter $F_1$ is used. 
    The red line shows the contour for which the sum of the two largest eigenvalues of the matrix $\boldsymbol{U}$ is 1.}
    \label{fig:apfig1}
\end{figure}

\item Case III: We have ($a_1=0$ and $b_1 \neq 0$) 

If $a_1=0$ and $b_1 \neq 0$, then from the Eq.\ref{ap2}), we see that we may need to suppress both 
$|00 \rangle \langle 00|$ and $|11 \rangle \langle 11|$. But local filters $F_1$ and $F_2$ can suppress only
one of them. Let us first use the filter $F_1$ and compute two largest eigenvalues of $\boldsymbol{U}$
for different values of $b_1$ and $h$ as shown in the Fig.~(\ref{fig:apfig2}). 

\begin{figure}[H]
    \centering
    \includegraphics[width=0.75\linewidth]{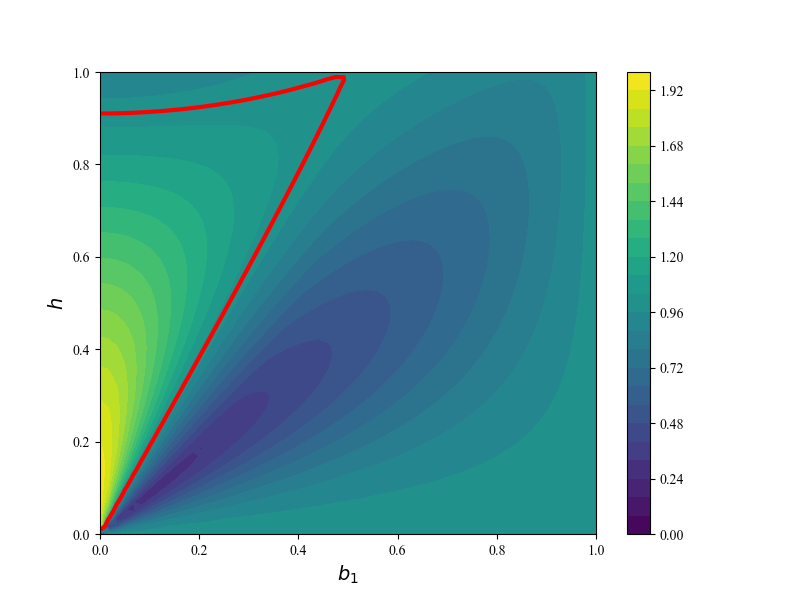}
    \caption{Contour plot of the sum of two largest eigenvalues of matrix $\boldsymbol{U}$ for $\Tilde{\rho}^{1}_{12}$,
    as a function of filter parameter $h$ and the state parameter $b_1$. Here, $a_1 = 0$ and the filter $F_1$ is used.
    The red line shows the contour for
    which the sum of two largest eigenvalues of matrix $\boldsymbol{U}$ is 1.}
    \label{fig:apfig2}
\end{figure}
We see Bell violation for small values of $b_1$, and the maximum range of Bell violation for the parameter $b_1$ is $0\leq b_1\leq0.5$, which is seen for $h=1$. We also see that with the increase in the strength of the filter, the violation range does not grow but shrinks. This might be because there is no clear separation of the noise from the entangled state.

Let us now use the filter $F_2$ and check the Bell violation for $\Tilde{\rho}^{2}_{12}$ by computing
the two largest eigenvalues of $\boldsymbol{U}$
for varying values of $b_1$ and $h$ as shown in the Fig.~(\ref{fig:apfig3}).


\begin{figure}[H]
    \centering
    \includegraphics[width=0.75\linewidth]{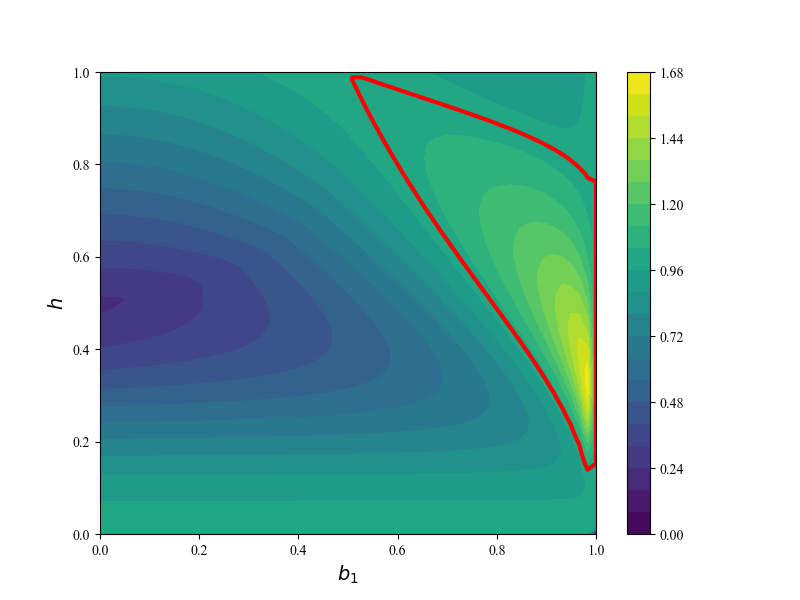}
    \caption{Contour plot of the sum of two largest eigenvalues of matrix $\boldsymbol{U}$ of $\Tilde{\rho}^{2}_{12}$,
    as a function of filter parameter $h$ and the state parameter $b_1$. Here, $a_1 = 0$ and the filter $F_2$ is used.
    The red line shows the contour for which the sum
    of two largest eigenvalues of matrix $\boldsymbol{U}$ is 1.}
    \label{fig:apfig3}
\end{figure}

We see the Bell violation for the large values of $b_1$ and the maximum range of Bell violation for the parameter $b_1$
is $b_1 \geq0.5$ which is seen for $h =1$. We also see that with the increase in the strength of the filter, the violation range does not grow but shrinks. This might be again because there is no clear separation of
the noise from the entangled state.  We thus see that the state in Eq.(\ref{ap1}) violates the Bell monogamy relation Eq.(\ref{eq4})
over the whole range of the parameter $b_1$.

\item  Case IV: The general case: ($a_1\neq$0 and $b_1\neq 0$)

We have seen in the previous cases that two types of filters, namely $F_1=h\ket{0}\bra{0}+\ket{1}\bra{1}$ and $F_2=\ket{0}\bra{0}+h\ket{1}\bra{1}$
do give us Bell violation for different ranges of variables $a_1$ and $b_1$ depending on the filter strength $h$. 
So we use filters $F_1$ and $F_2$ with various strengths of filters to check for the Bell violation.

Let us first use the filter $F_1$ and the corresponding filtered state $\Tilde{\rho}^{1}_{12}$ to find the range of parameters that gives the Bell violation.
We first fix the value of filter strength $h$ and then calculate the Bell violation numerically for various values of $a_1$ and $b_1$.
We then repeat the process by changing the value $h$. In the Fig.~(\ref{fig:apfig4}), we give an example of the
Bell violation for various ranges of $a_1$ and $b_1$ for filter strength $h=0.7$.
\begin{figure}[]
    \centering
    \includegraphics[width=0.75\linewidth]{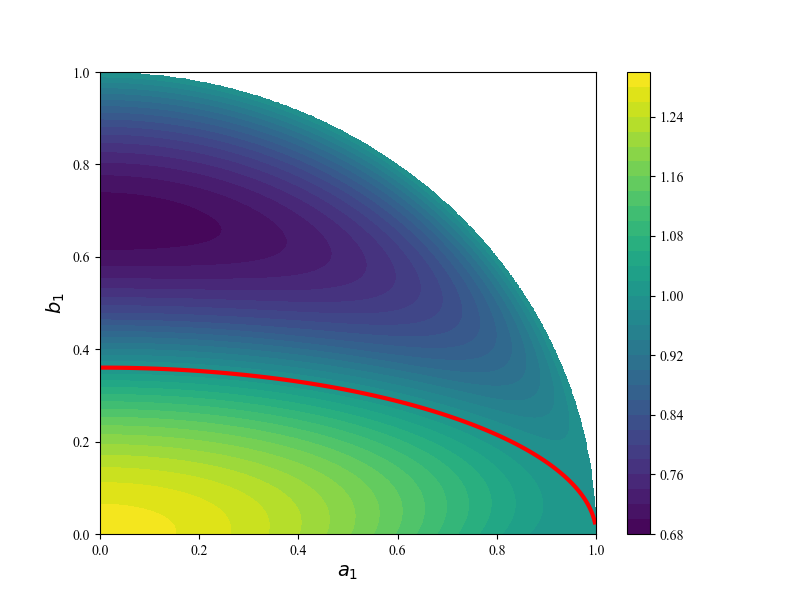}
    \caption{Contour plot of the sum of two largest eigenvalues of matrix $\boldsymbol{U}$ of $\Tilde{\rho}^{1}_{12}$
    as a function of $a_1$ and $b_1$. The filter strength $h=0.7$ and the filtering operation is $F_1$.}
    \label{fig:apfig4}
\end{figure}
We see that for a fixed value of $h$, the range of $a_1$ and $b_1$ for the Bell violation is a combination of the
previous two cases, {\it i.e.,} when $a_1 \neq 0, b_1=0$ and $a_1=0, b_1 \neq 0$. We see the violation for the whole range of $a_1$
and for small values of $b_1$. As the value of $a_1$ decreases, the range of Bell violation in $b_1$ increases and is maximum when $a_1=0.$

Next, we use the filter $F_2$ and the corresponding filtered state $\Tilde{\rho}^{2}_{12}$ to find the range of parameters that gives the Bell violation.
\begin{figure}[]
    \centering
    \includegraphics[width=0.75\linewidth]{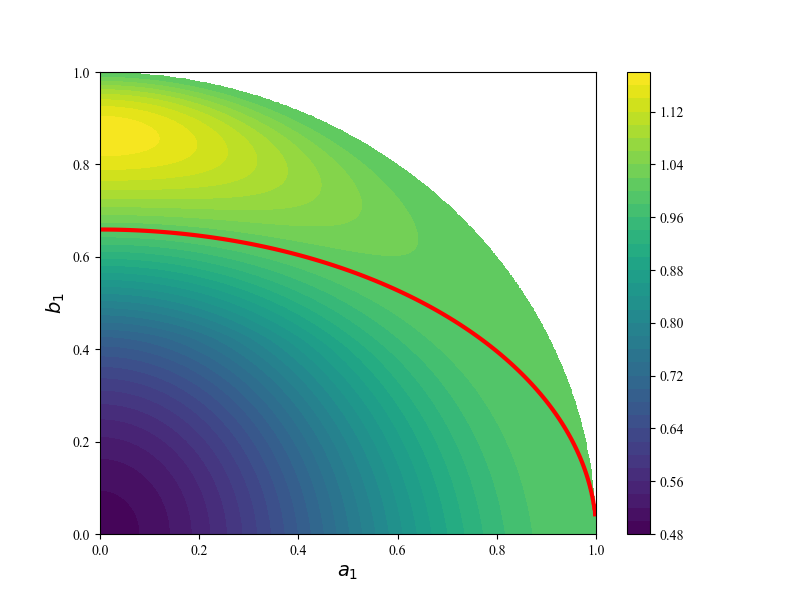}
    \caption{Contour plot of the sum of two largest eigenvalues of matrix $\boldsymbol{U}$ for $\Tilde{\rho}^{2}_{12}$ 
    as a function of $a_1$ and $b_1$. The filter strength $h=0.7$ and the filtering operation is $F_2$.}
    \label{fig:apfig5}
\end{figure}
We observe that for a fixed value of $h$, the range of $a_1$ and $b_1$ for Bell violation can be given as a
combination of the previous two cases, {\it i.e.} when $a_1 \neq 0 , b_1=0$ and $a_1=0, b_1 \neq 0$.
We see in the Fig.~(\ref{fig:apfig5}) the violation for almost the whole range of $a_1$ 
(constrained by $b_1, a_1^2+b_1^2=1$) and for the larger values of $b_1$. So again we see that after the local
filtering operations, the permutation-symmetric state in Eq.(\ref{ap1}) violates the Bell monogamy relation Eq.(\ref{eq4})
over most of the range of the parameters $a_1$ and $b_1$.
\end{itemize}\vspace{1cm}

 {\bf \Large Appendix B}
 \vspace{0.5cm}\\
In this appendix, we consider a more general three-qubit state and demonstrate the violation of Bell monogamy
relations. The most general three-qubit pure state can be written as \cite{acinstate}
\begin{equation}
    |\Psi\rangle=\lambda_0|000\rangle+\lambda_1 e^{i \varphi}|100\rangle+\lambda_2|101\rangle+\lambda_3|110\rangle+\lambda_4|111\rangle, \label{eq:acin}
\end{equation}
where $\lambda_i \geq 0$, $0\leq\varphi\leq\pi$ and $\sum_i\lambda_i^2=1$. Since the state $\ket{\Psi}$ is
not a permutation-symmetric state, there will be three different two-qubit reduced states, i.e., \begin{equation}
    \rho_{12}={\rm Tr_3}(\ket{\Psi}\bra{\Psi}), \quad  \rho_{13}={\rm Tr_2}(\ket{\Psi}\bra{\Psi}) \text{ and } \rho_{23}={\rm Tr_1}(\ket{\Psi}\bra{\Psi}).
\end{equation}
    Using the state in Eq.(\ref{eq:acin}), we can compute the density operators of the subsystems. To simplify the calculation,
we choose $\varphi =0, \lambda_0=0.1$, $\lambda_1=0$, $\lambda_2=0.2$ and $\lambda_3=0.3$. With this choice, we find that
 partial transposed density operator of each subsystem has a negative eigenvalue. Therefore, all three two-qubit subsystems
 are entangled. However, all three subsystems don't violate Bell-CHSH inequality, thus respecting the Bell
 monogamy relation Eq.(\ref{eq4}). We evaluate the matrix $\boldsymbol{U}$ for each subsystem to identify if any subsystem violates Bell's inequality. The sum of largest two eigenvalues of matrix $\boldsymbol{U}$ is found to be greater than 1 for subsystem $\rho_{23}$, and less than 1 for subsystem $\rho_{12}$ and $\rho_{13}$. So, only subsystem $\rho_{23}$ shows Bell violation. Other subsystems, though entangled, do not violate Bell-CHSH inequality. We now apply filtering operation to improve Bell violation in such subsystems.

The subsystem $\rho_{12}$ without filtering is
\begin{equation}
\rho_{12}=\left(
\begin{array}{cccc}
 0.01 & 0. & 0. & 0.03 \\
 0. & 0. & 0. & 0. \\
 0. & 0. & 0.04 & 0.185 \\
 0.03 & 0. & 0.185 & 0.95 \\
\end{array}
\right).
\end{equation}
To increase the sum of largest two eigenvalues of matrix $\boldsymbol{U}$, we apply filtering operation $F_1=\ket{0}\bra{0}+h\ket{1}\bra{1}$ on qubit 1 and $F_2=h\ket{0}\bra{0}+\ket{1}\bra{1}$ on qubit 2. The subsystem $\rho_{12}$ after filtering operation is
\begin{equation}
    \Tilde{\rho}_{12}=\frac{(F_1\otimes F_2)\rho_{12}(F_1^{\dagger}\otimes F_2^{\dagger})}{{\rm Tr}((F_1\otimes F_2)\rho_{12}(F_1^{\dagger}\otimes F_2^{\dagger}))}=\left(
\begin{array}{cccc}
 \frac{1}{4 h^2+96} & 0 & 0 & \frac{3}{4 h^2+96} \\
 0 & 0 & 0 & 0 \\
 0 & 0 & \frac{h^2}{h^2+24} & \frac{h\sqrt{43/2} }{h^2+24} \\
 \frac{3}{4 h^2+96} & 0 & \frac{h\sqrt{43/2} }{h^2+24} & \frac{95}{4 h^2+96} \\
\end{array}
\right).
\end{equation}
By setting the filtering parameter $h=0.4$, the sum of largest two eigenvalues of matrix $\boldsymbol{U}$ is 1.0012. Therefore, after filtering, the subsystem $\Tilde{\rho}_{12}$ violates Bell-CHSH inequality. 

Similarly, the subsystem $\rho_{13}$ without filtering is
\begin{equation}
\rho_{13}=\left(
\begin{array}{cccc}
 0.01 & 0. & 0. & 0.02 \\
 0. & 0. & 0. & 0. \\
 0. & 0. & 0.09 & 0.278 \\
 0.02 & 0. & 0.278 & 0.9 \\
\end{array}
\right).
\end{equation}
We apply filtering operation $F_1=\ket{0}\bra{0}+h\ket{1}\bra{1}$ on qubit 1 and $F_2=h\ket{0}\bra{0}+\ket{1}\bra{1}$ on qubit 2. The subsystem $\rho_{12}$ after filtering operation is
\begin{equation}
    \Tilde{\rho}_{13}=\frac{(F_1\otimes F_2)\rho_{13}(F_1^{\dagger}\otimes F_2^{\dagger})}{{\rm Tr}((F_1\otimes F_2)\rho_{13}(F_1^{\dagger}\otimes F_2^{\dagger}))}=\left(
\begin{array}{cccc}
 \frac{1}{9 h^2+91} & 0 & 0 & \frac{2}{9 h^2+91} \\
 0 & 0 & 0 & 0 \\
 0 & 0 & \frac{9 h^2}{9 h^2+91} & \frac{3 \sqrt{86} h}{9 h^2+91} \\
 \frac{2}{9 h^2+91} & 0 & \frac{3 \sqrt{86} h}{9 h^2+91} & \frac{90}{9 h^2+91} \\
\end{array}
\right).
\end{equation}
If we choose the filtering parameter $h=0.3$, the sum of largest two eigenvalues of matrix $\boldsymbol{U}$ is 1.0004. So, after filtering, the subsystem $\Tilde{\rho}_{13}$ also violates Bell-CHSH inequality. Since all three subsystems violate Bell-CHSH inequality, the Bell monogamy relation (\ref{eq4}) is violated.

\end{document}